\documentclass[%
 reprint,
superscriptaddress,
 amsmath,amssymb,
 aps,
floatfix,
]{revtex4-1}

\usepackage{graphicx}% Include figure files
\usepackage{dcolumn}% Align table columns on decimal point
\usepackage{bm}% bold math
\usepackage{hyperref}% add hypertext capabilities
\DeclareUnicodeCharacter{200E}{ }
\begin{document}

\preprint{APS/123-QED}

\title{Almost medium-free measurement of the Hoyle state direct-decay component with a TPC}% Force line breaks with \\
		\author{J.~Bishop}
		\affiliation{Cyclotron Institute, Texas A\&M University, College Station, TX 77843, USA}
		\affiliation{Department of Physics \& Astronomy, Texas A\&M University, College Station, TX 77843, USA}
		\author{G.V.~Rogachev}
		\affiliation{Cyclotron Institute, Texas A\&M University, College Station, TX 77843, USA}
		\affiliation{Department of Physics \& Astronomy, Texas A\&M University, College Station, TX 77843, USA}
		\affiliation{Nuclear Solutions Institute, Texas A\&M University, College Station, TX 77843, USA}
		\author{S.~Ahn}
		\affiliation{Cyclotron Institute, Texas A\&M University, College Station, TX 77843, USA}
		\author{E.~Aboud}
		\affiliation{Cyclotron Institute, Texas A\&M University, College Station, TX 77843, USA}
		\affiliation{Department of Physics \& Astronomy, Texas A\&M University, College Station, TX 77843, USA}
		\author{M.~Barbui}
		\affiliation{Cyclotron Institute, Texas A\&M University, College Station, TX 77843, USA}
		\author{A.~Bosh}
		\affiliation{Cyclotron Institute, Texas A\&M University, College Station, TX 77843, USA}
		\affiliation{Department of Physics \& Astronomy, Texas A\&M University, College Station, TX 77843, USA}
		\author{C.~Hunt}
		\affiliation{Cyclotron Institute, Texas A\&M University, College Station, TX 77843, USA}
		\affiliation{Department of Physics \& Astronomy, Texas A\&M University, College Station, TX 77843, USA}
		\author{H.~Jayatissa}
		\affiliation{Physics Division, Argonne National Laboratory, Argonne, IL 60439, USA}
		\author{E.~Koshchiy}
		\affiliation{Cyclotron Institute, Texas A\&M University, College Station, TX 77843, USA}
		\author{R.~Malecek}
		\affiliation{{Department of Physics and Astronomy, Louisiana State University, Baton Rouge, LA 70803, USA} }
		\author{S.T.~Marley}
		\affiliation{{Department of Physics and Astronomy, Louisiana State University, Baton Rouge, LA 70803, USA} }
		\author{E.C.~Pollacco}
		\affiliation{{IRFU, CEA, Universite Paris-Saclay, Gif-Sur-Ivette, France}}
		\author{C.D.~Pruitt}
		\affiliation{Department of Chemistry, Washington University, St. Louis, MO 63130, USA}
		\author{B.T.~Roeder}
		\affiliation{Cyclotron Institute, Texas A\&M University, College Station, TX 77843, USA}
		\author{A. Saastamoinen}
		\affiliation{Cyclotron Institute, Texas A\&M University, College Station, TX 77843, USA}
		\author{L.G.~Sobotka}
		\affiliation{Department of Chemistry, Washington University, St. Louis, MO 63130, USA}
		\author{S.~Upadhyayula}
		\affiliation{Cyclotron Institute, Texas A\&M University, College Station, TX 77843, USA}
		\affiliation{Department of Physics \& Astronomy, Texas A\&M University, College Station, TX 77843, USA}
	
\email{jackbishop@tamu.edu}

\date{\today}% It is always \today, today,

\begin{abstract}
\begin{description}
\item[Background]
The structure of the Hoyle state, a highly $\alpha$-clustered state at 7.65 MeV in $^{12}\mathrm{C}$, has long been the subject of debate. Understanding if the system comprises of three weakly-interacting $\alpha$-particles in the 0s orbital, known as an $\alpha$-condensate state, is possible by studying the decay branches of the Hoyle state.
\item[Purpose]
The direct decay of the Hoyle state into three $\alpha$-particles, rather than through the $^{8}\mathrm{Be}$ ground state, can be identified by studying the energy partition of the 3 $\alpha$-particles arising from the decay. This paper provides details on the break-up mechanism of the Hoyle stating using a new experimental technique.
\item[Method]
By using beta-delayed charged-particle spectroscopy of $^{12}\mathrm{N}$ using the TexAT (Texas Active Target) TPC, a high-sensitivity measurement of the direct 3 $\alpha$ decay ratio can be performed without contributions from pile-up events.
\item[Results]
A Bayesian approach to understanding the contribution of the direct components via a likelihood function shows that the direct component is $<0.043\%$ at the 95\% confidence level (C.L.). This value is in agreement with several other studies and here we can demonstrate that a small non-sequential component with a decay fraction of about $10^{-4}$ is most likely.
\item[Conclusion]
The measurement of the non-sequential component of the Hoyle state decay is performed in an almost medium-free reaction for the first time. The derived upper-limit is in agreement with previous studies and demonstrates sensitivity to the absolute branching ratio. Further experimental studies would need to be combined with robust microscopic theoretical understanding of the decay dynamics to provide additional insight into the idea of the Hoyle state as an $\alpha$-condensate. 
\end{description}
\end{abstract}

\pacs{Valid PACS appear here}% PACS, the Physics and Astronomy
\maketitle

%\tableofcontents
%\section{\label{sec:Introduction}Introduction}
\emph{I. Introduction.}
Near-threshold states in $^{12}\mathrm{C}$ have a large effect on the formation of elements. Through the triple-alpha process, the synthesis bottleneck associated with the instabilities of the A=5 and 8 isobars is overcome. This reaction is enhanced by several orders of magnitude by the existence of a $0^{+}$ state just above the 3$\alpha$ threshold known as the Hoyle state. The structure of the state has been an area of interest since its discovery \cite{Hoyledisc}. While the fact that the Hoyle state is a highly-clustered 3$\alpha$ structure is common knowledge, the exact nature of the clustering is a subject of debate to this date, and has ramifications for other light few-body systems involved in nucleosynthesis. It has been suggested that the Hoyle state may be the manifestation of a new state of matter known as an $\alpha$-condensate \cite{THSR}.
When the average nuclear density drops below 1/3 of its normal value, the lowest-energy state is a bosonic cluster of $\alpha$-particles, a state with some properties similar to a Bose-Einstein condensate. Such a hypothesis has received extensive study in the past decade theoretically, but experimental observables for such an exotic state are extremely difficult to obtain \cite{HoyleFamily}. One relevant observable is the direct 3-body decay of the Hoyle state, i.e. bypassing the $^{8}\mathrm{Be}$(g.s) intermediate. If an $\alpha$-condensate were to exist, this branching ratio can be predicted using a simple formulation of the $\alpha$-condensate wavefunction in conjunction with two- and three-body tunneling calculations. This value is very small and is estimated at 0.06 \% \cite{Smith,DDP2} although extracting a value is highly model dependent. The latest high-sensitivity experiments \cite{Smith,DellAquila,Rana} can only provide upper limits for this value and the best limit currently lies at 0.019\% \cite{Rana}. Beyond this point, one reaches the limitations of background associated with the use of silicon detector arrays \cite{Robinthesis}.  Recent indirect methods predict a branching ratio of 0.00057$\%$ \cite{RobinPRC}, a factor of 45 lower than the current limit, indicating the magnitude of the improvement likely needed to directly measure the direct decay channel.\par
One may differentiate between the decay mechanisms, sequential and direct, by the energy-partition of the 3 $\alpha$-particles. The sequential decay mechanism restricts the energy of one of the $\alpha$-particles (in the center-of-mass frame) to roughly 50\% due to the well-constrained momentum and energy conservation associated with $^{12}\mathrm{C} \rightarrow \alpha + {^{8}\mathrm{Be}}$. Direct decays have no such energy restriction and can occupy the full available phase space for three-body decays. The most-likely direct decay components correspond to when all $\alpha$-particles share the energy equally.
\par
\emph{II. Experimental setup}
To study the role of the direct decay to the Hoyle state decays, excited states in $^{12}\mathrm{C}$ were populated using the $\beta$-delayed charged-particle spectroscopy technique \cite{NIMJEB} using the TexAT (Texas Active Target) TPC (Time Projection Chamber) \cite{TexATNIM}. A $^{12}\mathrm{N}$ beam was produced using the K500 cyclotron at the Cyclotron Institute at Texas A\&M University. This beam was created via the interaction of an 11 MeV/nucleon $^{10}\mathrm{B}$ primary beam undergoing a $^{3}\mathrm{He}(^{10}\mathrm{B},{^{12}\mathrm{N}})n$ reaction in a gas cell. The beam of interest was then selected using MARS (Momentum Achromatic Recoil Spectrometer) \cite{MARS} and delivered into the TexAT. TexAT is a general-purpose TPC using Micromegas (MICRO MEsh GASeous) + THick Gas Electron Multipliers (THGEM) amplification and segmentation. The signals induced on the Micromegas are digitized at 10 MHz by the GET (General Electronics for TPCs) \cite{GET} and written to disk. This experiment was performed in the `2p-mode' made available by GET whereby two half-events are taken to disk. The first half of the event corresponds to the implanting of the $^{12}\mathrm{N}$ into TexAT. The second half corresponds to the decay of $^{12}\mathrm{C}$ into 3 $\alpha$-particles. For decays that proceed via the $^{12}\mathrm{C}$ ground-state or first-excited state, the second half of the event is absent but the partial half event is still taken to disk. As discussed in further detail for this experimental setup, this allows for one-at-a-time implant and decay spectroscopy using 20 Torr CO$_2$. Details of the experimental setup and analysis of the data are provided in depth in Ref.~\cite{NIMJEB}.

\par

\emph{III. Almost medium-free branching ratio measurements of the Hoyle state decay.}
Unlike observables extracted from heavy-ion reactions, the use of $\beta$ decay to populate the Hoyle state provides direct access to an almost medium-free determination of a direct 3-body decay. Furthermore, this route takes maximal advantage of the characteristics of TPCs to remove the contributions from pile-up events and other effects that contribute to the limit currently achieved using solid-state arrays. The intrinsic limitations for identifying different decays inside a TPC correspond to low-energy scattering of the particles in the fill gas and limited segmentation/thresholds which influence the accuracy of track reconstruction of the decay-particles. In order to identify any rare direct decays in the data set, each track was fitted with three arms, one for each decay $\alpha$-particle. The initial parameters for these arms were seeded by a Hough transform \cite{Hough} and the decay vertex was identified by a combination of using the stopping point of the implanting $^{12}\mathrm{N}$ beam and the highest energy deposition point of the decay tracks. Due to scattering effects in the gas, these tracks may deviate from their original momentum vector introducing an uncertainty in the measured final momentum vector. In order to minimize this uncertainty, we employed a technique that ensures exact momentum conservation between the three $\alpha$-particles \cite{NIMJEB, funkifit}. As a consequence, the uncertainty in the length of the longest track was reduced and the ability to identify direct decay improved. Our procedure to identify direct decays makes use of two experimental parameters, one using two extracted angles and the other making use of standard Dalitz plots. These are described, and shown below. \par
\begin{figure}
\centerline{\includegraphics[width=0.45\textwidth]{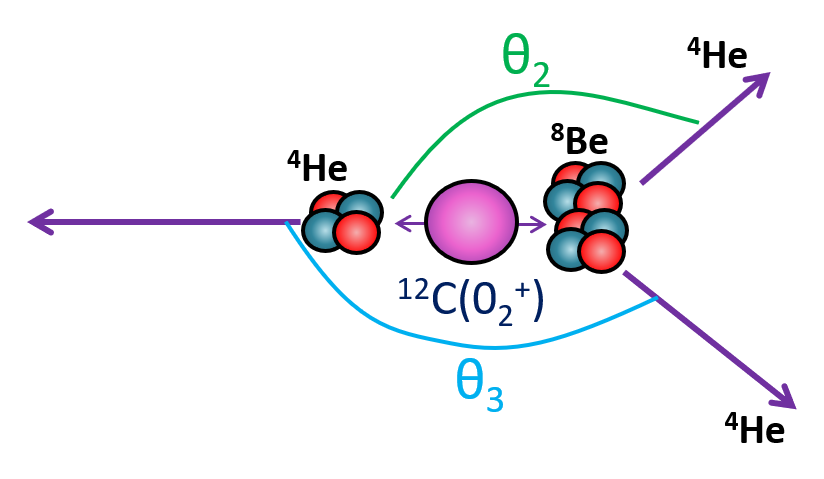}}
\caption{Definition of $\theta_{2}$ and $\theta_{3}$ as the angle between the longest $\alpha$-particle track and the second and third-longest respectively.\label{fig:deftheta2theta3}}
\end{figure}
\begin{figure}
\centerline{\includegraphics[width=0.5\textwidth]{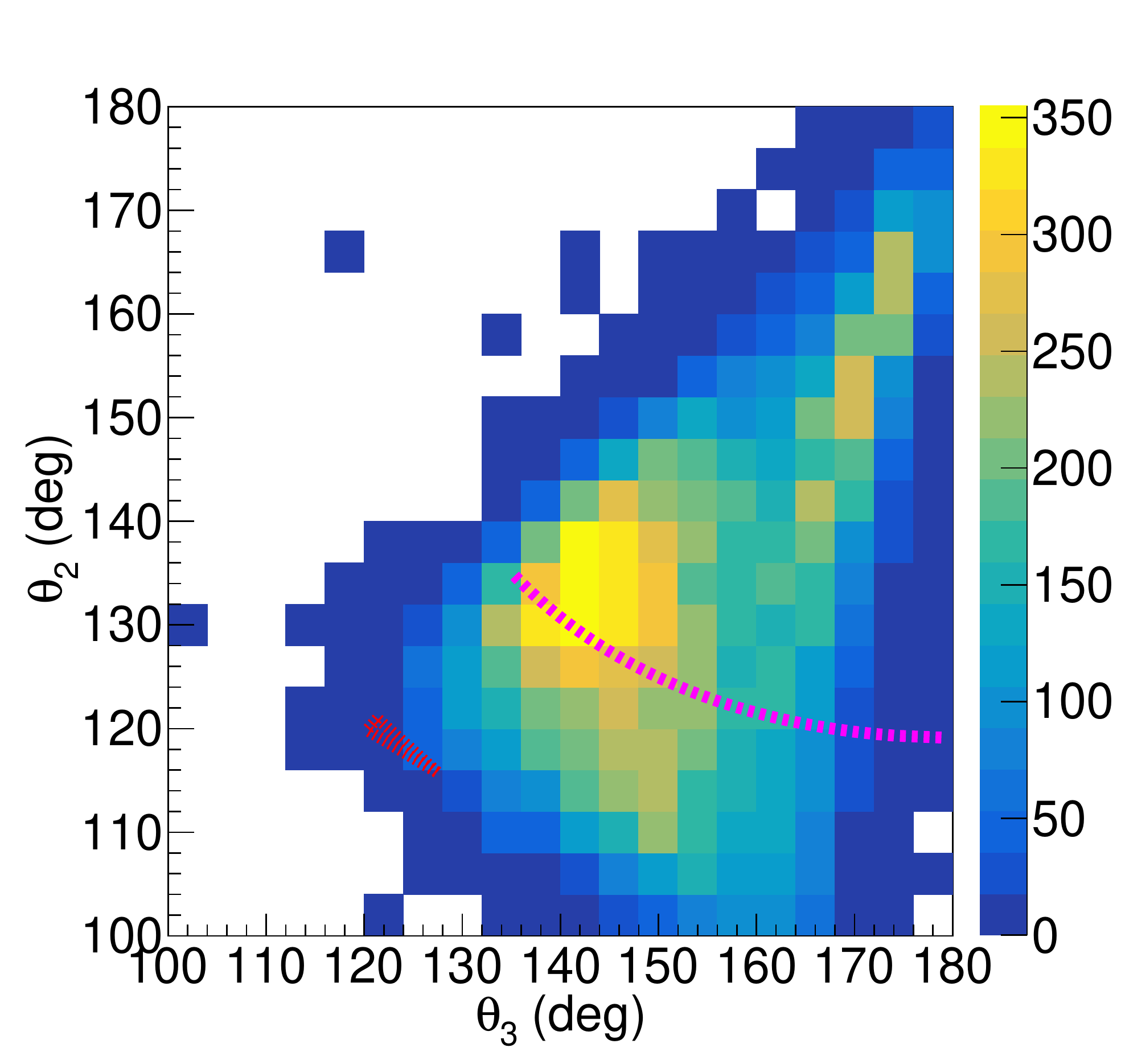}}
\caption{Reconstructed $\theta_{2}$ against $\theta_{3}$ after kinematic fitting. The locus for sequential decay is shown with the dashed magenta line. The region occupied by direct decays is shown by a dash-dotted red line and is focused mainly around $120^{\circ},120^{\circ}$ for both $\theta_2 ,\theta_3$.\label{fig:theta2theta3}}
\end{figure}

\emph{a. Angular decay information}
The angles between the most-energetic $\alpha$-particle and the two others (as shown in Fig.~\ref{fig:deftheta2theta3}), are determined by a fitting procedure to the 3 $\alpha$-particle tracks. The results (after kinematic fitting) from the data are shown in Fig.~\ref{fig:theta2theta3}, overlaid with the locus for sequential and direct decay. The events for direct decay would be centered on ($120^{\circ},120^{\circ})$ for an equal-energy partition. 
While the dominance of sequential decay is clear, additional information is required for clear identificiation of any direct decays. 
\par
\begin{figure}
\centerline{\includegraphics[width=0.5\textwidth]{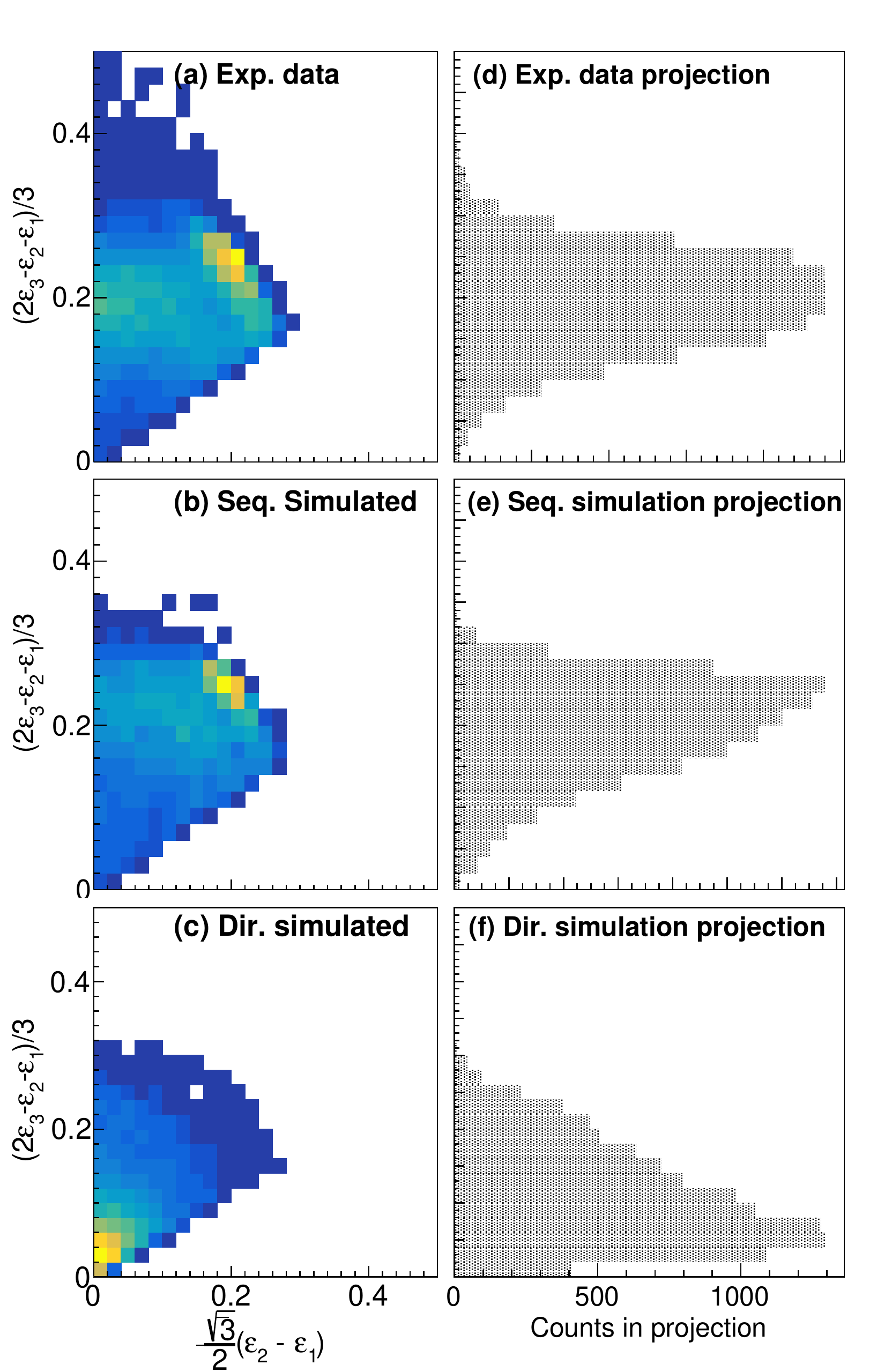}}

\caption{Dalitz plots for (a) experimental data, (b) simulated fully-sequential decays and (c) simulated fully-direct (DDP$^2$) decays. The projections of these Dalitz plots onto the y-axis are shown in (c), (d) and (e) for the experimental data, simulated sequential and simulated direct decays respectively.\label{fig:Dalitz}}
\end{figure}

\emph{b. Dalitz formulation}
The Dalitz plot affords a convenient way to show the population of the available phase space in three-body decays. By taking linear combinations of the partial $\alpha$-particle energies in the center-of-mass frame, $\varepsilon_i$, such that $\sum_i^3 \varepsilon_i = 1$, the energy partition of the 3 $\alpha$-particles can be represented on a 2-dimensional plot. Figure~\ref{fig:Dalitz} shows how the linear combinations of these three parameters can differentiate sequential and direct decay. Here, the direct decay component is simulated using the DDP$^2$  (Direct Decay Phase space + Penetrability) model \cite{DDP2} that weights an otherwise uniform population of phase space by the three-body penetrabilities. The Dalitz population within this model reconstructed with TexAT is shown in Fig.~\ref{fig:Dalitz}c. This preferentially populates the center of the Dalitz plot where all the $\alpha$-particles have similar energies. It is therefore practically identical to the DDE (Direct Decay Energy-sharing) model where the $\alpha$-particles have identical energies, smeared only by the uncertainty principle \cite{HoyleFamily,DDP2}. The experimental data, shown in Fig.~\ref{fig:Dalitz} demonstrate the dominance of the sequential decay (as per Fig.~\ref{fig:Dalitz}b).
\begin{figure}
\centerline{\includegraphics[width=0.5\textwidth]{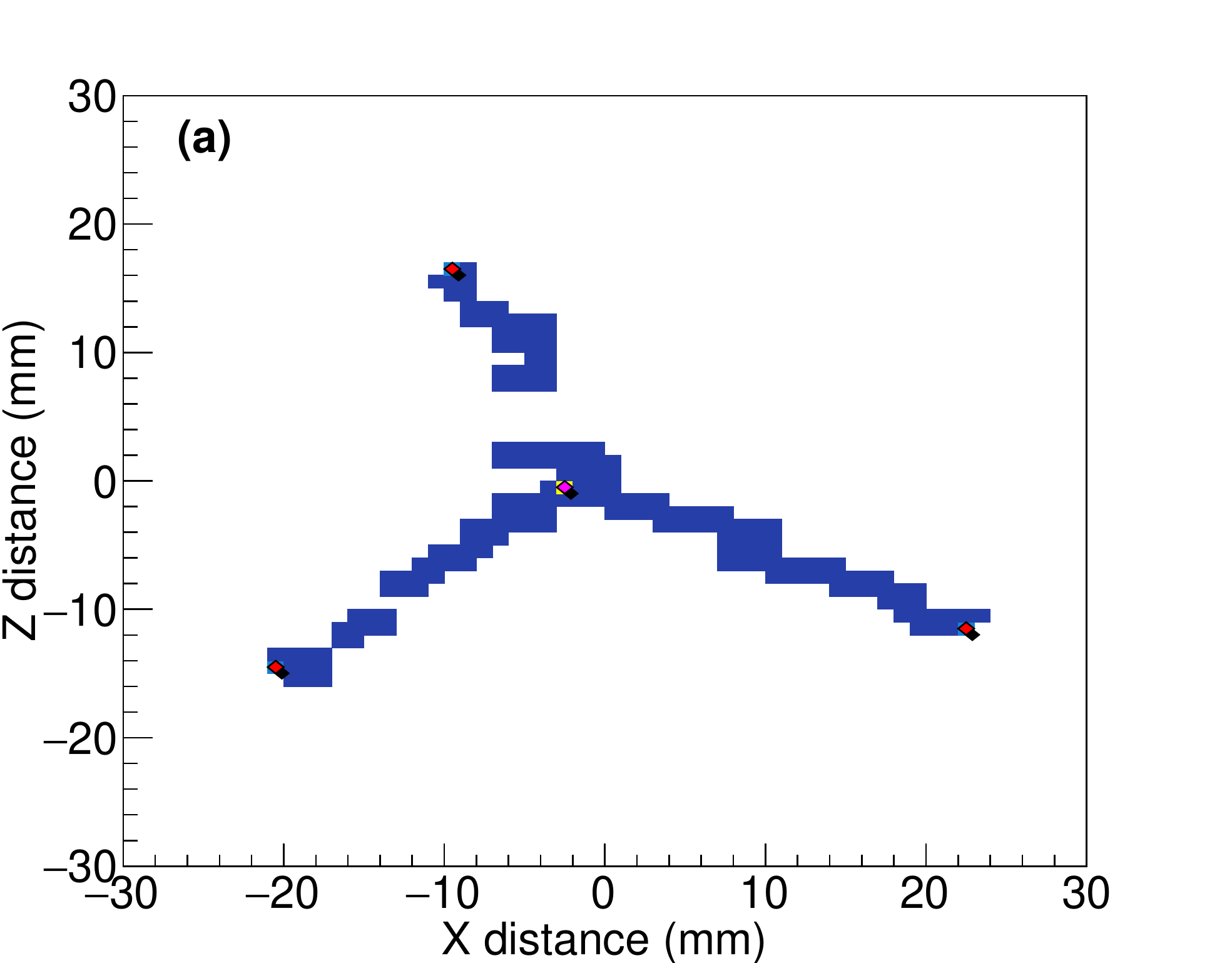}}
\centerline{\includegraphics[width=0.5\textwidth]{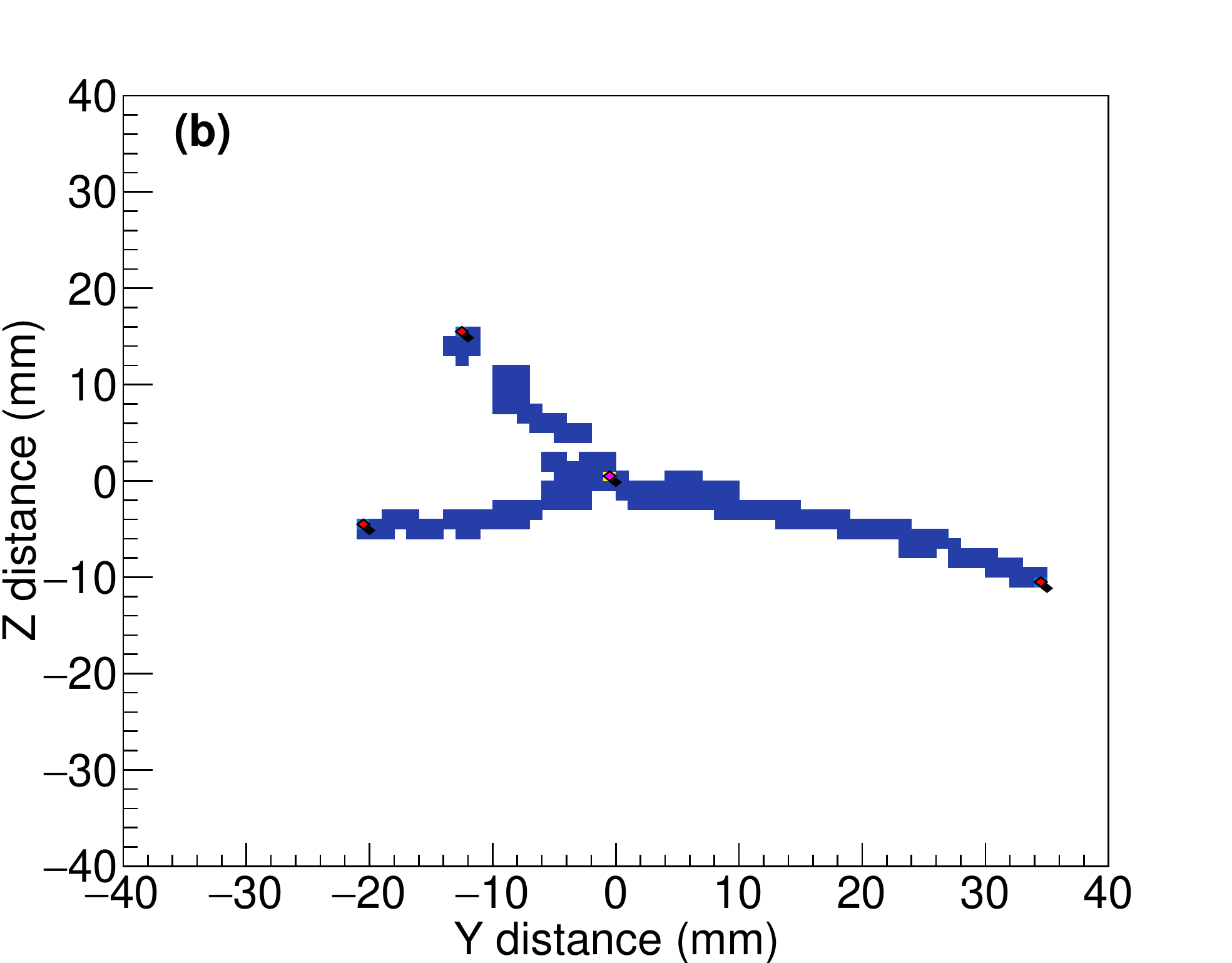}}
\caption{(a) An example of a direct-looking event looking at a side-on (XZ) projection where the beam is implanted along the y-axis (into the plane of the plot) and the z-axis corresponds to the drift axis. The three $\alpha$-particle arms have very similar lengths/energies ($\varepsilon = 0.37, 0.35, 0.28$) and the $\theta_{23}$ angles between the $\alpha$-particles are $119^{\circ}$ and $130^{\circ}$. The combination of these parameters favors the classification of this event as direct-looking and has a p-value for direct/sequential decay of 0.1 and $1.5 \times 10^{-5}$ respectively. (b) An example of a sequential event looking at a side-on (YZ) projection. The p-values for direct/sequential decay are $1.9 \times 10^{-7}$ and $0.6$ respectively. The decay vertex and extremes of the three arms are marked by magenta and red diamonds respectively for both events (color online). \label{fig:directevent}}
\end{figure}
\par
\emph{c. Branching ratio measurement}
Taking the angular information and location on the Dalitz plot, the $\chi^2$ was evaluated for each event for both sequential and direct decay (incorporating the varying contribution of the direct component across the Dalitz plot). This was formulated as follows:
\begin{equation}
\chi^2_{\theta} = \frac{\text{min} \{(\theta_2-\theta_{2_{\text{theory}}})^2 + (\theta_3-\theta_{3_{\text{theory}}})^2 \}}{\sigma^2_{\theta}}, 
\end{equation}
where $\theta_{i_{\text{theory}}}$ is determined for either the direct or sequential case and $\sigma_{\theta}$ is the experimental error determined via the Monte Carlo simulation ($5^{\circ}$).  The kinematics for sequential decay constrain $\theta_{2},\theta_{3}$ to the locus as shown in Fig.~\ref{fig:theta2theta3}. For direct decay, the equal $\alpha$-particle energy constraint is slightly relaxed so that the highest-energy $\alpha$-particle fractional energy cannot exceed $\varepsilon=0.35$ which generates a small region around $\theta_{2},\theta_{3} = 120^{\circ},120^{\circ}$, shown in Fig.~\ref{fig:theta2theta3} by the dash-dotted red line. As with the sequential case, the shortest distance to this locus is found. This $\chi^2_{\theta}$ value was also combined with $\chi^2_{D}$ from the Dalitz plot measurement, defined as:
\begin{equation}
\chi^2_{D} = 
\begin{cases}
(\frac{y-y_{\text{seq}}}{\sigma_D})^2, \text{ for sequential}\\
(\frac{x^2+y^2}{\sigma_D^2}), \text{ for direct}
\end{cases}
%\frac{1}{\sigma_{\theta}} \frac{1}{\sigma_{\text{Dalitz}}}
\end{equation}
where $x$ and $y$ are the Dalitz plot co-ordinates for each event, $y_{\text{seq}}$ is the expected Dalitz co-ordinate for sequential decay and $\sigma_D$ (=0.059) is the experimental error determined via the width of the projection of the experimental data shown in Fig.~\ref{fig:Dalitz}d. The expected $y_{\text{seq}}$ is $\sim \frac{1}{6}$ however the experimentally-observed value is slightly offset at 0.2 which is used for the $\chi^2$ formulation. This offset is also replicated in the GEANT4 simulations (Fig.~\ref{fig:Dalitz}e) and is attributed to a combination of energy-loss uncertainties at the low energies causing a slight systematic shift as well as threshold effects inside the TPC which cannot be fully corrected for. This effect is dominant for small $\alpha$-particle energies and, as such, does not greatly affect the center of the Dalitz plot where the $\alpha$-particles have a sufficiently large energy ($\sim 130$ keV each). The global $\chi^2$ is then defined as the sum of $\chi^2_{\theta}$ and $\chi^2_{D}$.\par
A total of 19019 Hoyle decay events were taken as the cleanest unbiased subset of data whereby the implanted beam stops sufficiently centrally in the TexAT sensitive area such that no $\alpha$-particle may escape and such that the beam stops in the central region of the Micromegas where the detector has no multiplexing \cite{TexATNIM}. Therefore, the decay vertex can be much more confidently identified, thereby improving the fitting and energy-partition determination. By selecting the most direct-looking events (where $\chi^2_{dir}~<~\chi^{2}_{seq}$) and manually checking these 224 events had the decay vertex and corresponding decay arms correctly identified, a double-check was possible to ensure that sequential events were not erroneously misidentified as direct decays. A small subset of events (9), after manual checks, still had $\chi^2$-values that indicated that the event was more direct-looking than sequential-looking and corresponded to good Hoyle-state decays. An example event of which is shown in Fig.~\ref{fig:directevent}a in contrast to an example sequential event in Fig.~\ref{fig:directevent}b. Due to the finite resolution afforded by small-angle scattering and longitudinal straggling effects, it may still be that these events are statistically outlying sequential decays rather than direct decays. To determine the relative probabilities, the $\chi^2$ values were converted into p-values, $p_{\chi}$. These describes the probabilities that if the event was either direct or sequential, it would produce the observed values. For direct decays, the intensity distribution of the DPP$^2$ is applied at this point. \\
\begin{figure}
\centerline{\includegraphics[width=0.5\textwidth]{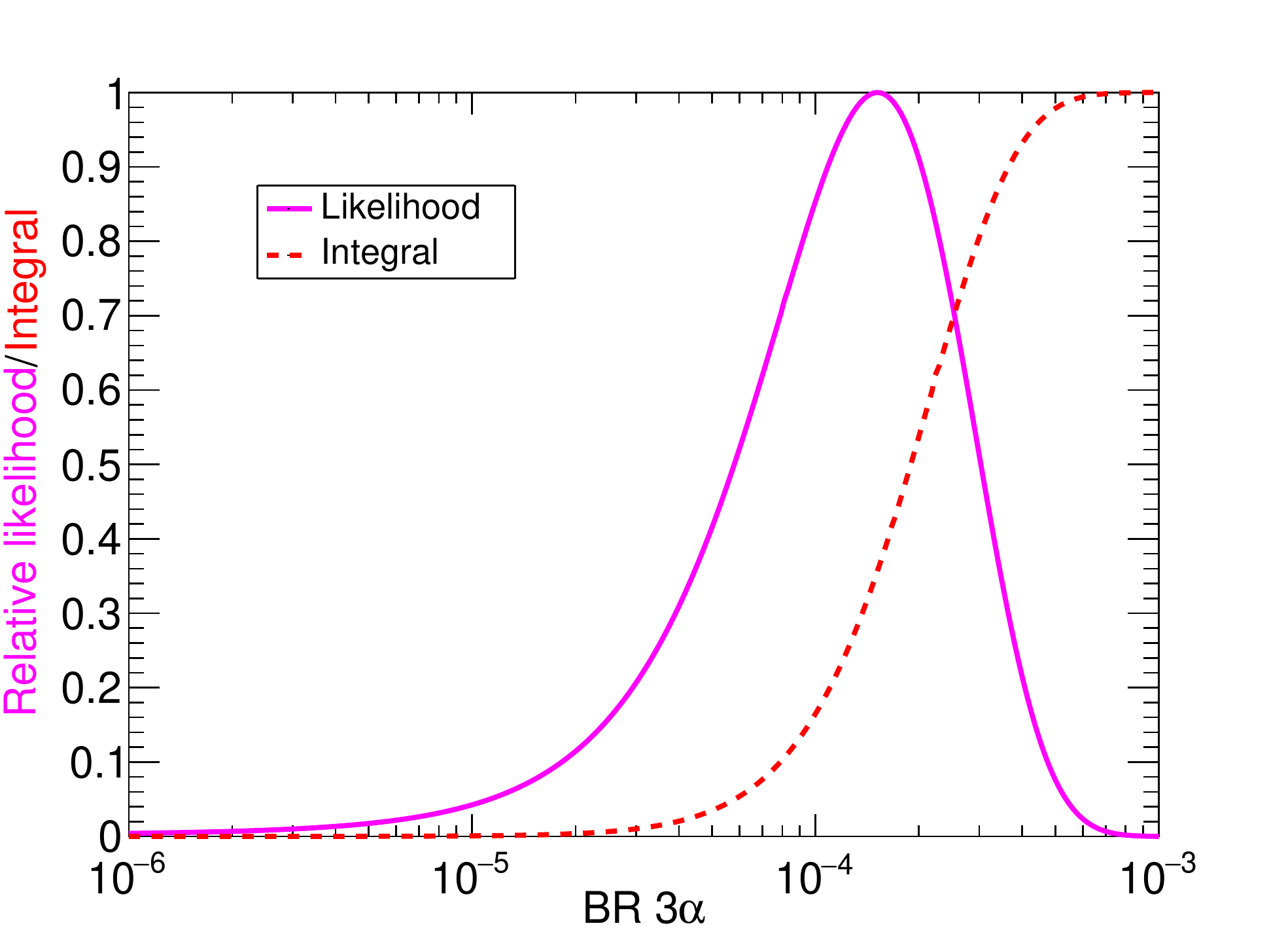}}
\caption{(Solid magenta) Likelihood function for different values of the direct 3$\alpha$ branching ratio ($\delta$) for our data using the formulation in Eq.~\ref{eq:likelihood}. (Dashed red) Integral of the relative likelihood function. At the 95\% C.L., the branching ratio is $<0.043\%$.\label{fig:likelihood}}
\end{figure}
The probability that an event is sequential is given via:
\begin{align}
P_{seq} = p_{\chi_{seq}} (1-\delta),
\end{align}
where $\delta$ is the direct 3$\alpha$ branching ratio. Similarly, the probability this event is direct is given by:
\begin{align}
P_{dir} = p_{\chi_{dir}} \delta.
\end{align}
The event-by-event probabilities were then used to create a log-likelihood distribution as a function of the 3$\alpha$ branching ratio:
\begin{align}
\mathcal{L}(\delta) = \sum_n \log({p_{\chi_{seq}} (1-\delta) + p_{\chi_{dir}} \delta}), \label{eq:likelihood}
\end{align}
which is representative of the product of the probability of each event being direct or sequential. This was then used to generate the likelihood function and the 95\% C.L. can be set from the integral of this likelihood function. These plots are shown in Fig.~\ref{fig:likelihood}. At the 95\% C.L., one can determine that the 3$\alpha$ branching ratio is $<0.043\%$. \par
Similarly, one may also use the likelihood function to set a lower limit for the direct decay. As mentioned above, there are a small number of good candidate direct decay events that require a reasonable branching ratio in order to explain their presence that cannot be explained by being a solely sequential decay that is a statistical outlier. From our data, this lower branching ratio is $>0.0058\%$ at 95\% C.L.
\par
The overall upper-limit is driven mainly by counting statistics with the obtained likelihood function matching well the Poisson distribution with $\lambda  = 0.017\%$ around the peak of the likelihood function. As such, one requires an increase in the statistics of a factor of 4 for a two-fold reduction in the upper limit ($\sim$ 80,000 counts). The influence of straggling in the gas and other finite-resolution effects are not the dominant contribution to the likelihood function as the overall separation between direct and sequential events is sufficient – this is apparent when looking at the ratio of the probabilities for sequential and direct decays for direct-looking events (as shown in Fig.\ref{fig:directevent}a).\par
%The contribution from the $^{8}\mathrm{Be}$ ghost-peak as examined by Refsgaard \cite{Refsgaard} can have a large contribution to events in the center of the Dalitz plot. As such, one cannot feasibly disentangle truly direct decays with these events from the $^{8}\mathrm{Be}$ ghost peak as depending on the model and parameters used, this may be practically identical. In addition to confirming the previous measurements of the upper-limit branching ratio, this current result is therefore very useful in determining the contribution of this ghost peak as the few direct-looking events can be said with certainty to \textbf{not} be due to pileup which was seen as an issue in previous experiments using silicon detectors.
It can be seen that the preferred (most likely) branching ratio is 0.01$\%$ and, given the use of a TPC in this experiment, one can be sure that this small number of events corresponds to real 3$\alpha$ decays, being a combination of real direct-decays and contributions from the so-called `ghost peak' in $^{8}\mathrm{Be}$ \cite{Refsgaard}. This ghost peak appears when one has a near-threshold resonance when the penetrability factor rises faster than the steeply-dropping but still long-tailed form (i.e. Breit-Wigner) of resonance line-shapes \cite{Barker}. A non-zero branching ratio was also predicted in previous studies \cite{Smith,HoyleFamily} although the ability to determine these as direct-looking rather than pileup on an event-by-event basis was not possible.
\par
\emph{IV. Conclusion}
An almost medium-free measurement of the Hoyle direct decay to three $\alpha$-particles has been performed with a TPC. With 95\% C.L., the direct branching ratio is $<0.043\%$. Contributions to the direct-looking events may correspond to a sizeable contribution from the $^{8}\mathrm{Be}$ ghost peak \cite{Refsgaard} and analyzing these events show that they are genuine 3$\alpha$-decays and are not pileup events as experienced with previous experiments that measured a similar upper-limit. The preferential $10^{-4}$ branching ratio seen here is in agreement with predictions from Faddeev calculations \cite{Ishikawa}. The strength of this work relies on the use of a TPC and removal of uncertainties related to pile-up, a problem that plagued all previous measurements. More sensitive experimental studies of the direct component of the Hoyle state decay will also require a better theoretical understanding of the 3$\alpha$-particle dynamics at the microscopic level in general and the contribution of the sequential decays to the $^{8}\mathrm{Be}$ ghost state in particular. \par
\emph{V. Acknowledgments}
This work was supported by the U.S. Department of Energy, Office of Science, Office of Nuclear Science, under award no. DE-FG02-93ER40773 and by National Nuclear Security
Administration through the Center for Excellence in Nuclear Training and University Based Research (CENTAUR) under grant number DE-NA0003841. G.V.R. also acknowledges the support of the Nuclear Solutions Institute.
\bibliography{HoyleBib}% Produces the bibliography via BibTeX.

\end{document}